\documentclass[aps,prd,twocolumn,groupedaddress]{revtex4-1}
\usepackage[dvips]{graphicx}
\usepackage{dcolumn}
\usepackage{bm} 
\usepackage{epsfig}
\usepackage{amsfonts}
\usepackage{amssymb,amscd}
\usepackage{subfigure}
\usepackage{xcolor}
\usepackage{verbatim}

\newcommand{\Pelotas}{High and Medium Energy Group, Instituto de F\'isica e Matem\'atica,
             Universidade Federal de Pelotas\\
             Caixa Postal 354,  96010-900, Pelotas, RS, Brazil.}

\newcommand{\FURG}{Instituto de Matem\'atica Estat\'istica e F\'isica, Universidade Federal de Rio Grande \\ Av. Italia, km 8, Campus Carreiros, CEP 96203-900, Rio Grande, RS, Brazil.}

\begin{document}
\title{Diffractive double quarkonium production at the LHC}
\author{C. Brenner Mariotto$^{1}$, V. P. Gon\c calves$^{2}$, R. Palota da Silva$^{2}$}
\affiliation{$^{1}$ \FURG\\ $^{2}$ \Pelotas}

\begin{abstract}
In this paper we study the inclusive and diffractive double quarkonium production. Based on the nonrelativistic QCD (NRQCD) factorization formalism for the quarkonium production mechanism we estimate the rapidity and transverse momentum dependence of the cross section for the $J/\Psi J/\Psi$ and $\Upsilon \Upsilon$ production in diffractive processes at LHC energies. The contributions of the color-singlet and color-octet channels are estimated and predictions for the total cross sections in the kinematical regions of the LHC experiments are also presented. Our results demonstrate that the contribution of diffractive processes is not negligible and that its study can be useful to test the Resolved Pomeron model.
\end{abstract}


\maketitle

\section{Introduction}
In a double quarkonium production a pair of mesons, e.g. $J/\Psi J/\Psi$ or $\Upsilon \Upsilon$, is produced in an independent hadronic colision \cite{Sun,Brambilla,Landsberg}. This subject is of great interest since it makes it possible to investigate the models that explain the process. At the LHC, due to the high density of partons in the initial state, the probability of a pair of mesons being produced in an independent collision must be significantly large. Recent observations of double $J/\Psi$ production were made by LHCb \cite{Aaij 1,Aaij 2}, which allows the comparison with theoretical predictions.

In a $pp$ collision one may have inclusive and exclusive processes. The class of exclusive events include diffractive reactions. Diffractive processes are those with no quantum numbers exchange, where it is possible to observe the presence of large, non asymptotically suppressed, rapidity gaps separating the centrally produced particles from the outgoing protons \cite{Bjorken}. This type of reaction, in opposition to the inclusive process, presents a very clean signal in the final state which makes it possible to analyze physical processes free of additional radiation contamination. It is believed that a high energy diffractive reaction is mediated by a Pomeron exchange, which is assumed to have a partonic structure \cite{Ingelman}.

In this contribution we investigate the inclusive and diffractive double quarkonium production in $pp$ collisions at the Run 2 LHC energies. For the diffractive reaction we study the single and double diffractive processes considering the Resolved Pomeron Model \cite{Ingelman}. This model assumes that the hard processes take place in a Pomeron-proton or a Pomeron-Pomeron interaction. Based on the nonrelativistic QCD (NRQCD) factorization formalism for the quarkonium production mechanism \cite{Bodwin} we estimate the rapidity and transverse momentum dependencies of the cross sections for the $J/\Psi J/\Psi$ and $\Upsilon \Upsilon$ production. The contributions of the color-singlet \cite{Qiao} and color-octet \cite{Ko} channels are taken into account and predictions for the total cross sections in the kinematical regions of the LHC experiments are also presented.  Our goal in this work is to study the heavy quarkonium production as a complementary analysis of diffractive processes and of the Pomeron structure. In addition we compare the magnitude of the inclusive cross section with the single and double diffractive ones (see \cite{BrennerMariotto:2018eef} for a more complete study).

\section{Diffractive double quarkonium production and NRQCD}

The single (double) diffractive process is characterized by the presence of one (two) intact forward hadron and one (two) rapidity gap(s), as well by a central object produced by the hard interaction and by soft particles, associated to the Pomeron remnants which are characteristic of the resolved Pomeron model.

\begin{figure*}
\begin{tabular}{ccc}
\centerline{
{
\includegraphics[scale =.5]{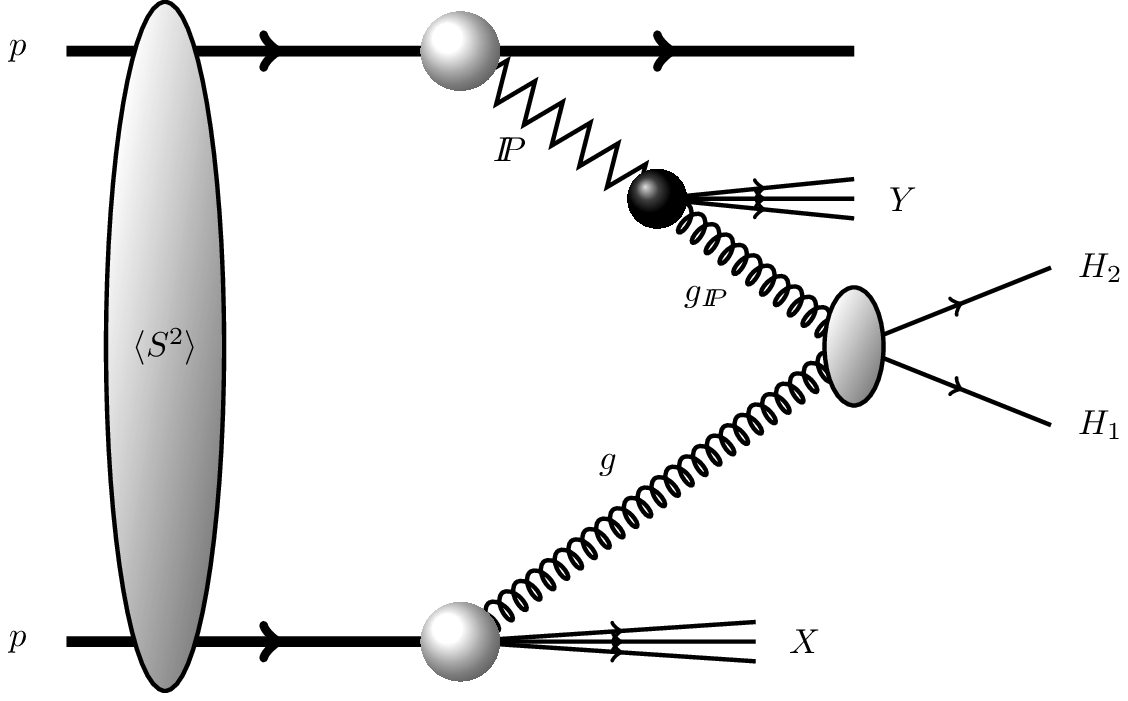}
}
{
\includegraphics[scale =.5]{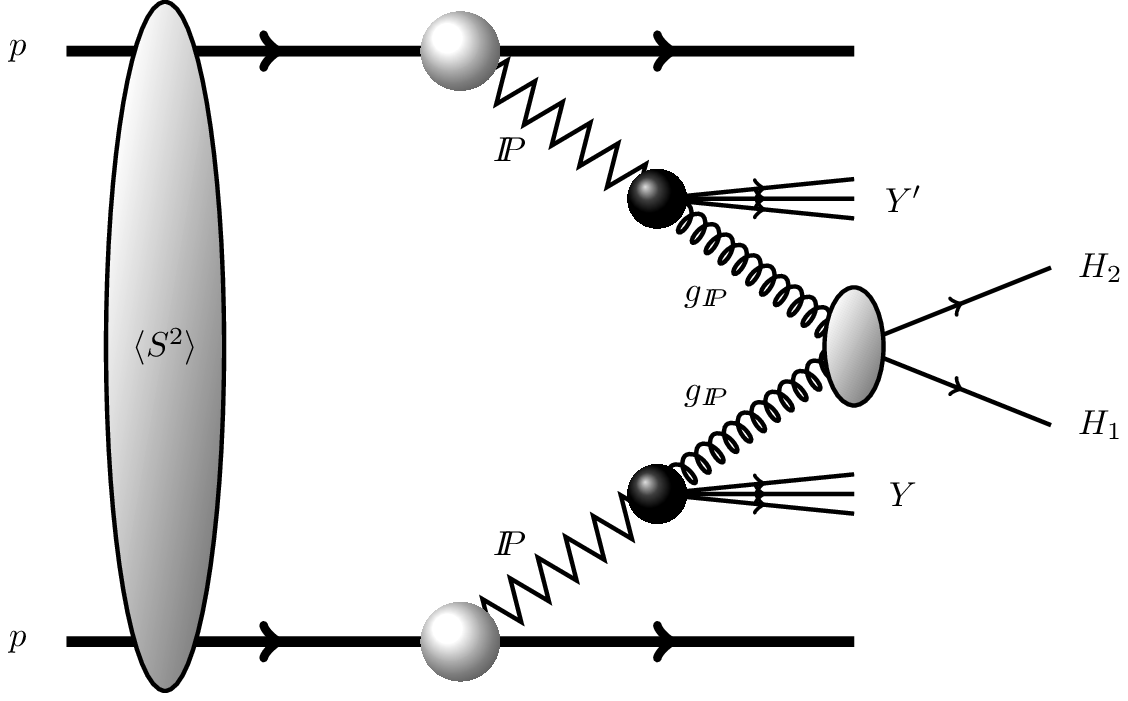}
}
{
\includegraphics[scale=.5]{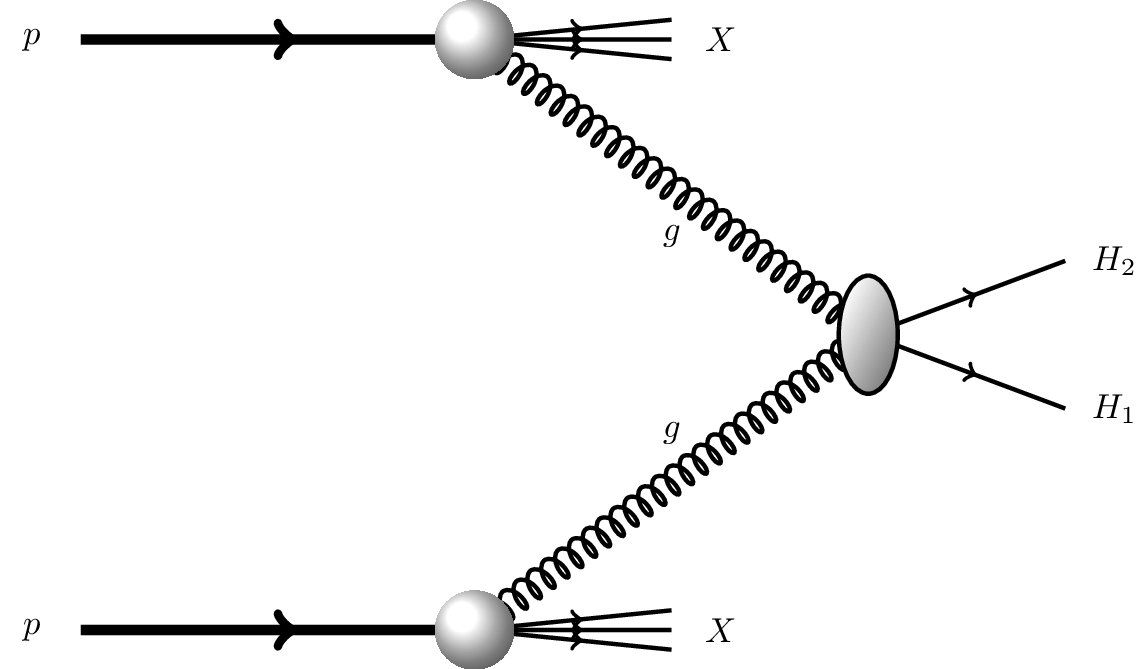}
}
}
\end{tabular}
\caption{Diagrams representing double quarkonium production in single diffractive (SD) (left panel), double diffractive (DD) (middle panel) and  inclusive processes (right panel.} 
\label{Diagramas}
\end{figure*}

In the NRQCD formalism \cite{Bodwin}, the cross section for the central diffractive production of a heavy quarkonium pair can be written as
\begin{widetext}
\begin{eqnarray}
d\sigma = \langle S^2 \rangle\sum_{a,b,n_1,n_2}f_{a/p}^D\otimes f_{b/p}^D\otimes d\hat{\sigma}[ab\to Q\bar{Q}[n_1]+Q\bar{Q}[n_2]+X]\cdot\langle O_{n_1}^{V1}\rangle\langle O_{n_2}^{V2}\rangle
\label{Geral}
\end{eqnarray}
\end{widetext}
where $\langle S^2 \rangle$ is the gap survival probability and $\otimes$ represents the presence of a rapidity gap in the final state, $f^D_{a/p}$ and $f^D_{b/p}$ are the diffractive parton distributions from the colliding protons, and the coefficients $d\hat{\sigma}[ab\to Q\bar{Q}[n_1]+Q\bar{Q}[n_2]+X]$
are perturbatively calculable short distance cross sections for the production of the two heavy quark pairs in an intermediate Fock state $n_i={}^{2S+1}L_J^{[i]}$ ($i=1,8$), which does not have to be color neutral. The $\langle {\cal{O}}^{V_i}_{n_i}\rangle$
are nonperturbative long-distance matrix elements (LDME), which describe the transition of the intermediate $Q\bar{Q}$ in the physical state $V$ via soft gluon radiation.
In the Color Singlet Model \cite{csm}, only the states with the same quantum numbers as the resulting quarkonium contribute to the formation of a bound $Q\bar{Q}$ state. In contrast, in NRQCD, also color octet $Q\bar{Q}$ states contribute to the quarkonium production cross section via soft gluon radiation. For Single Diffractive processes one have to replace one of the $f^D$ for the standard inclusive PDF's, $f_p$, whereas for inclusive processes one has two standard inclusive PDF's. In our calculations we will consider the fit B of the H1 parametrization for the diffractive PDFs. Moreover, we assume that $\langle S^2 \rangle = 0.02$ (0.05) for DD (SD) processes.

At high energies the double quarkonium production is dominated by gluon - gluon interactions. In the figure \ref{Diagramas} we present a generic diagram for the production of a pair of heavy vector mesons in Single Diffractive (SD) (left), Double Diffractive (DD) (middle) and inclusive (right) processes. In the SD diagram one has a Pomeron-proton interaction, whereas in the DD diagram one has a Pomeron-Pomeron interaction. The presence of a Pomeron is responsible for one rapidity gap and one outgoing proton in the final state. The second Pomeron in the DD diagram doubles the number of rapidity gaps and intact outgoing protons in respect to the SD process. It is also necessary to consider the probability that particles resulting from the underlying events are not produced. This probability is represented by the gap survival factor, $\langle S^2 \rangle$.

The differential cross section for double quarkonium production ($H_1=H_2$) can be written as
\begin{widetext}
\begin{eqnarray}
\frac{d\sigma }{dydp_T^2}=  \int_{x_{a\, min}} dx_a
{g^D(x_a,\mu^2)g^D(x_b,\mu^2)}\frac{x_ax_b}{2x_a-\bar{x}_Te^y}
\sum_{i=1,8} \frac{d\hat{\sigma}}{d\hat{t}}[gg\rightarrow
2 Q\bar{Q}_i(^3S_1)]
\cdot \langle {\cal{O}}_i^{V}(^3S_1) \rangle^2  \,\,,
\label{csdif2}
\end{eqnarray}
\end{widetext}
where 
$x_{a\, min}=\frac{\bar{x}_Te^{y}}{2-\bar{x}_Te^{-y} }$, 
$x_b=\frac{x_a\bar{x}_Te^{-y}}{2x_a-\bar{x}_Te^y}$, 
$\bar{x}_T=\frac{2m_T}{\sqrt{s}}$ and $m_T=\sqrt{M^2+p_T^2}$. Here  $M$ is the quarkonium mass, $p_T$ its transverse momentum and $y$ its rapidity. The quarkonium transverse mass is taken as the hard scale of the problem, with $\mu_R=\mu_F=m_T$.
Here, ${g^D(x_i,\mu^2)}$  are the diffractive gluon distribution functions from the two colliding protons (for the inclusive gluon PDF we use CTEQ6L \cite{cteq}). Finally, $\frac{d\hat{\sigma}}{d\hat{t}}$ in Eq. (\ref{csdif2}) are the hard scattering differential cross sections. The Feynman diagrams can be classified into two groups (Ref. \cite{Ko}): (a) diagrams associated to the nonfragmentation contribution (31 Feynman diagrams), with the leading contribution being the color singlet $(Q\bar{Q})_1({}^3S_1) + (Q\bar{Q})_1({}^3S_1)$ channel, and (b) diagrams associated to the gluon fragmentation contribution (72 Feynman diagrams), with the main contribution associated to the color octet $(Q\bar{Q})_8({}^3S_1) + (Q\bar{Q})_8({}^3S_1)$ channel. 

\begin{figure*}[t]
\begin{tabular}{c c}
\centerline{
{
\includegraphics[scale = .4]{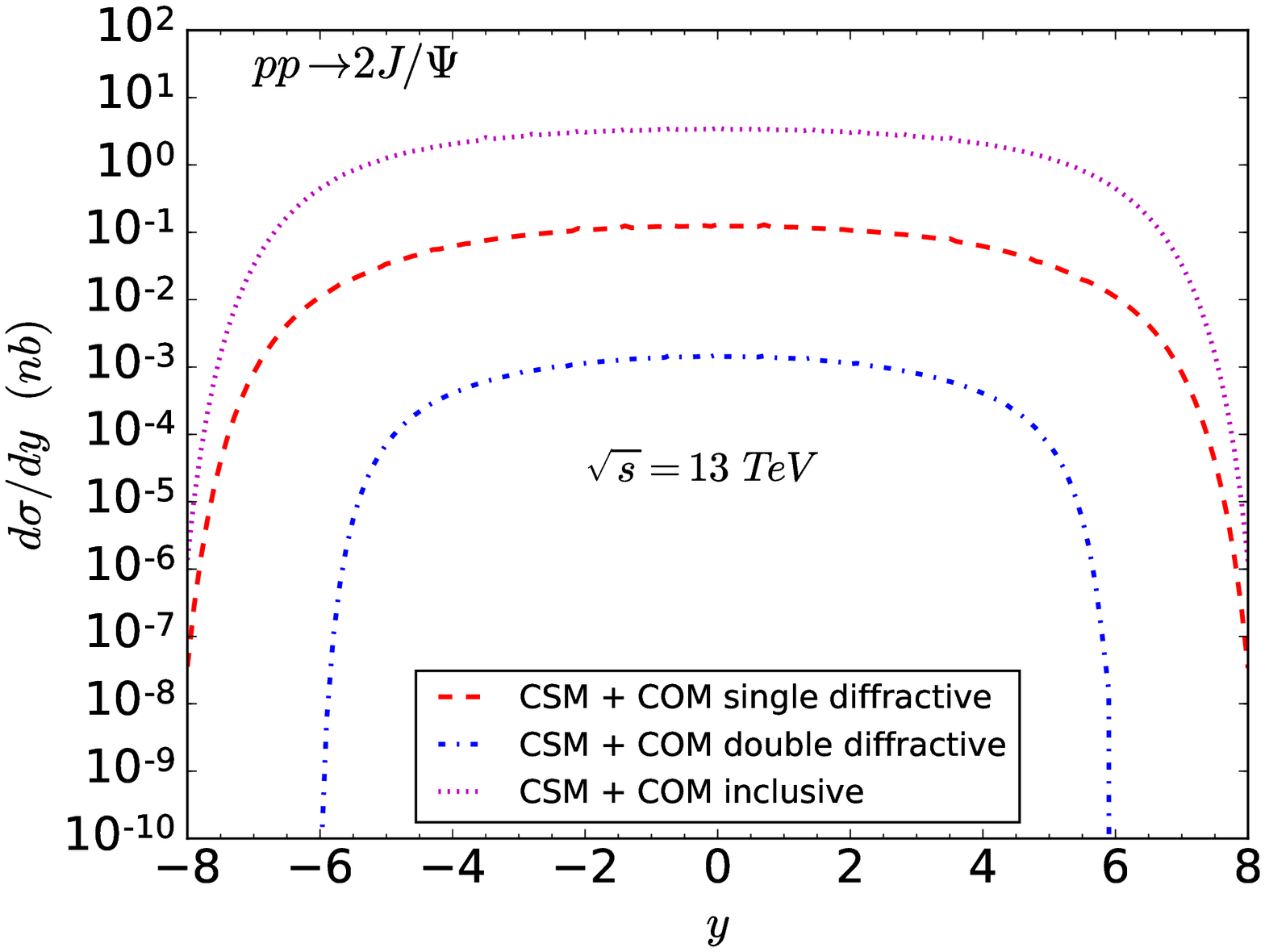}
}
{
\includegraphics[scale = .4]{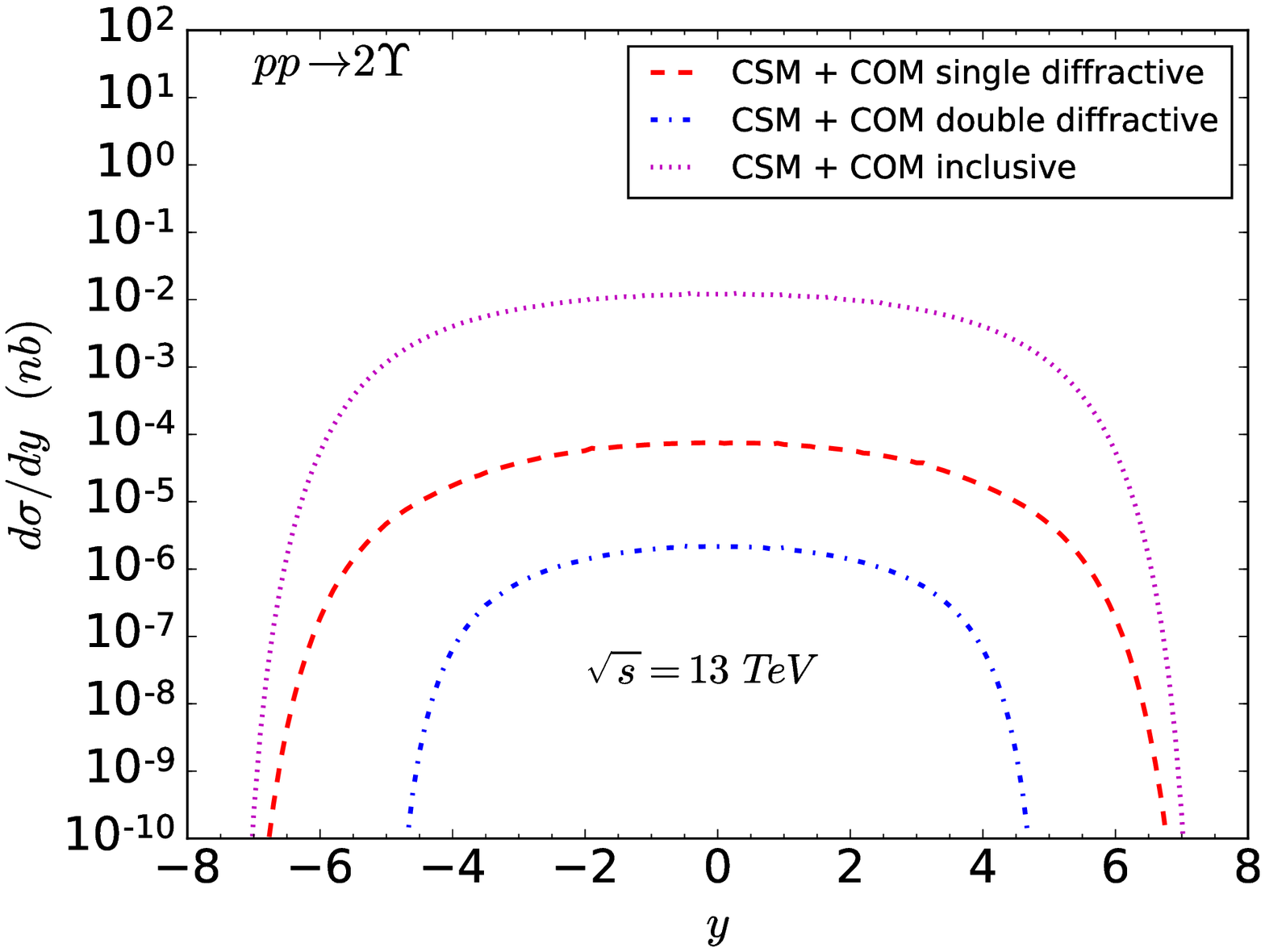}
}}
\end{tabular}
\caption{Rapidity distribution for double $J/\Psi$ production (left) and double $\Upsilon$ production (right).} 
\label{Rapidez}
\end{figure*}
\begin{figure*}
\begin{tabular}{c c}
\centerline{
{
\includegraphics[scale = .4]{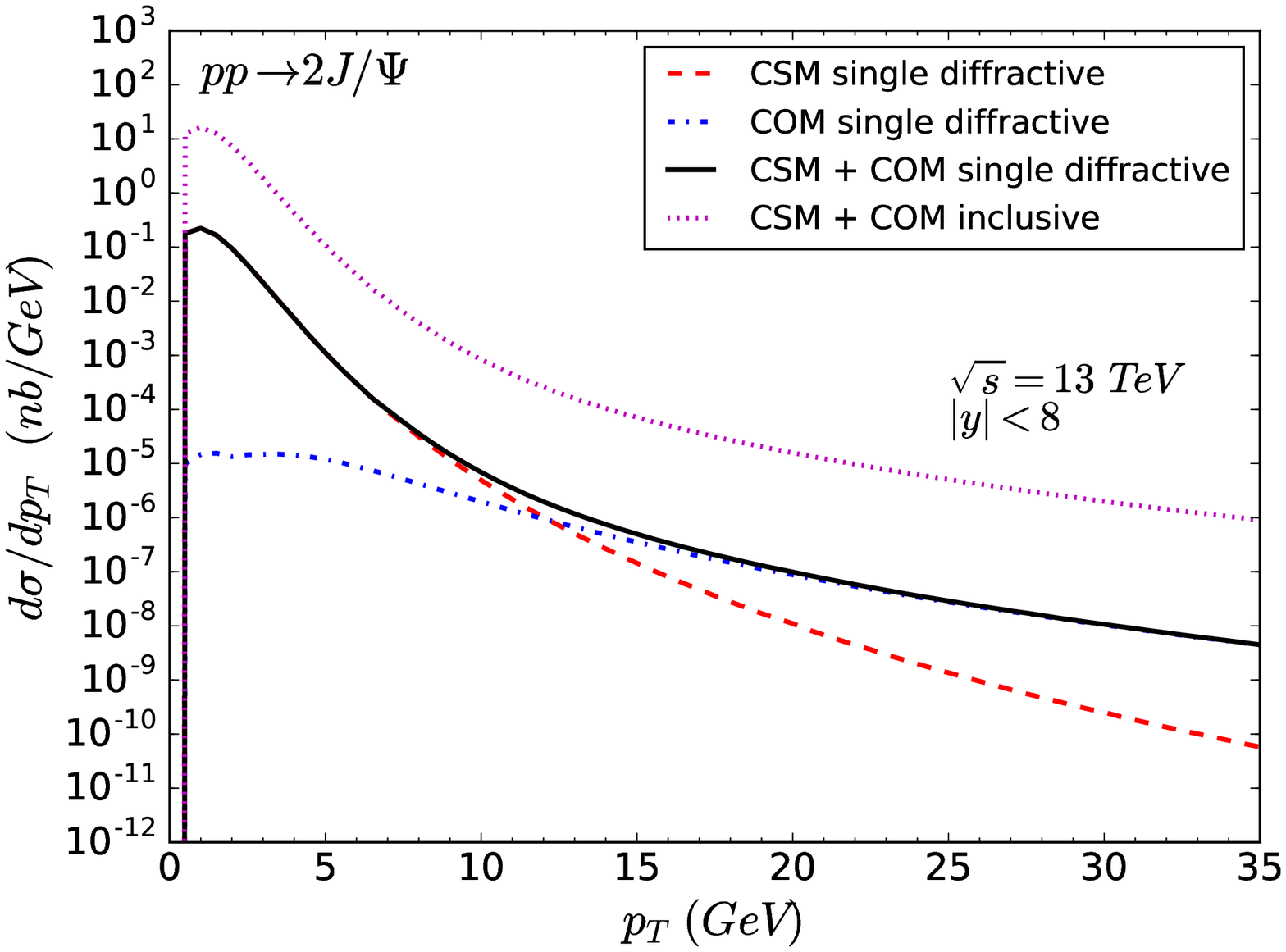}
}
{
\includegraphics[scale = .4]{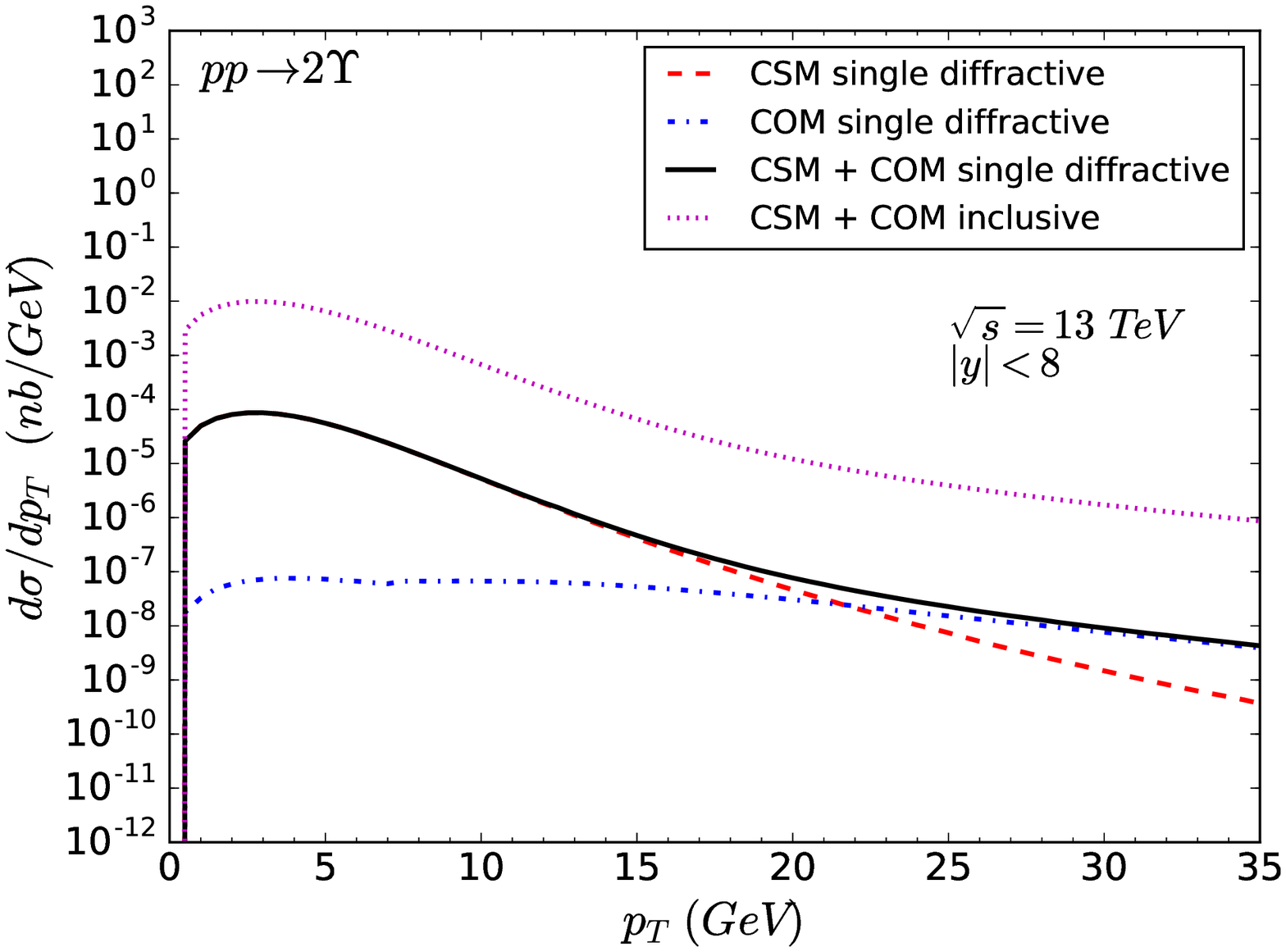}
}}
\end{tabular}
\caption{ $p_T$ distribution for double $J/\Psi$ production (left) and double $\Upsilon$ production (right) in single diffractive process.} 
\label{SD}
\end{figure*}

\begin{figure*}
\begin{tabular}{c c}
\centerline{
{
\includegraphics[scale = .4]{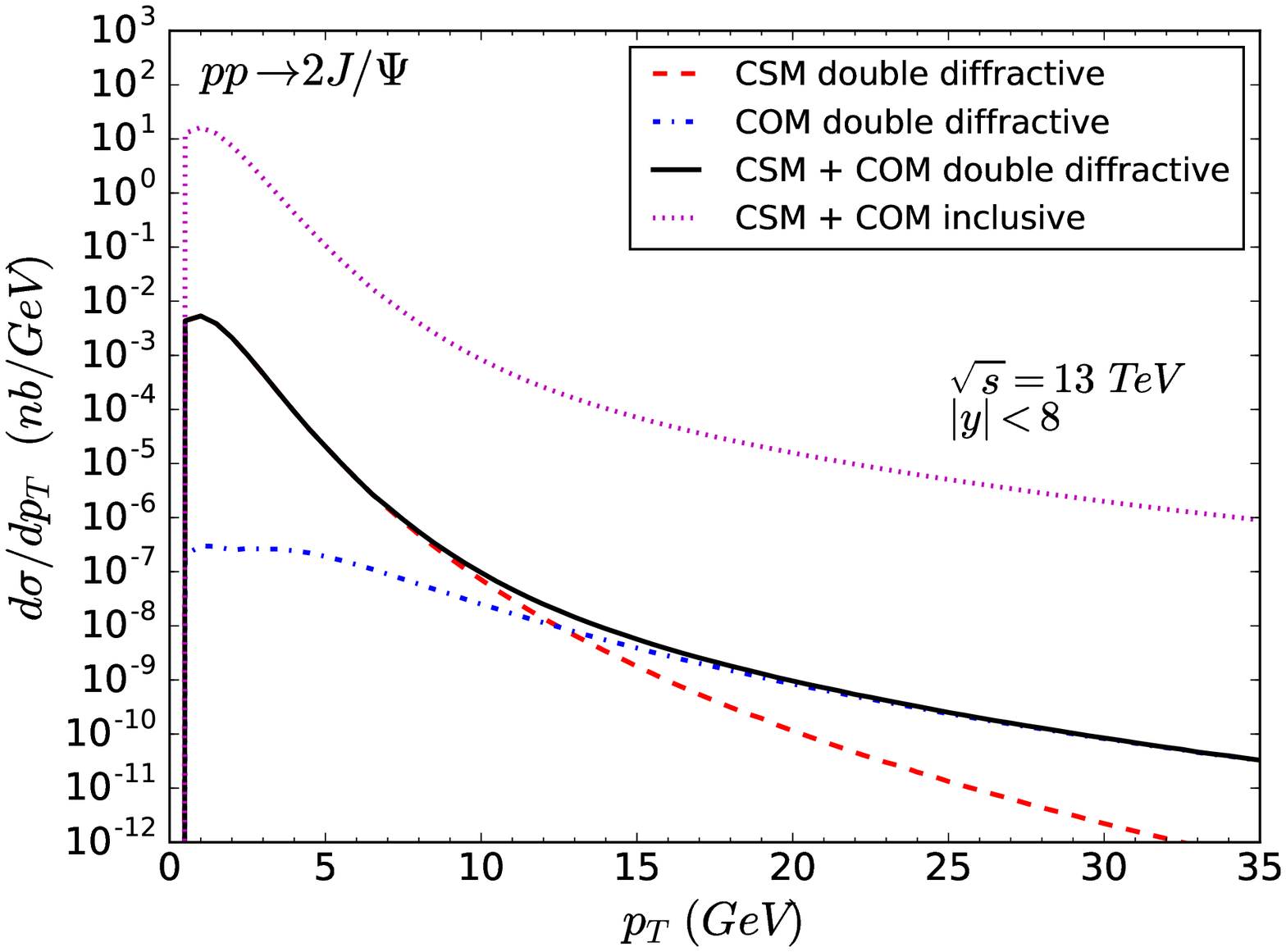}
}
{
\includegraphics[scale = .4]{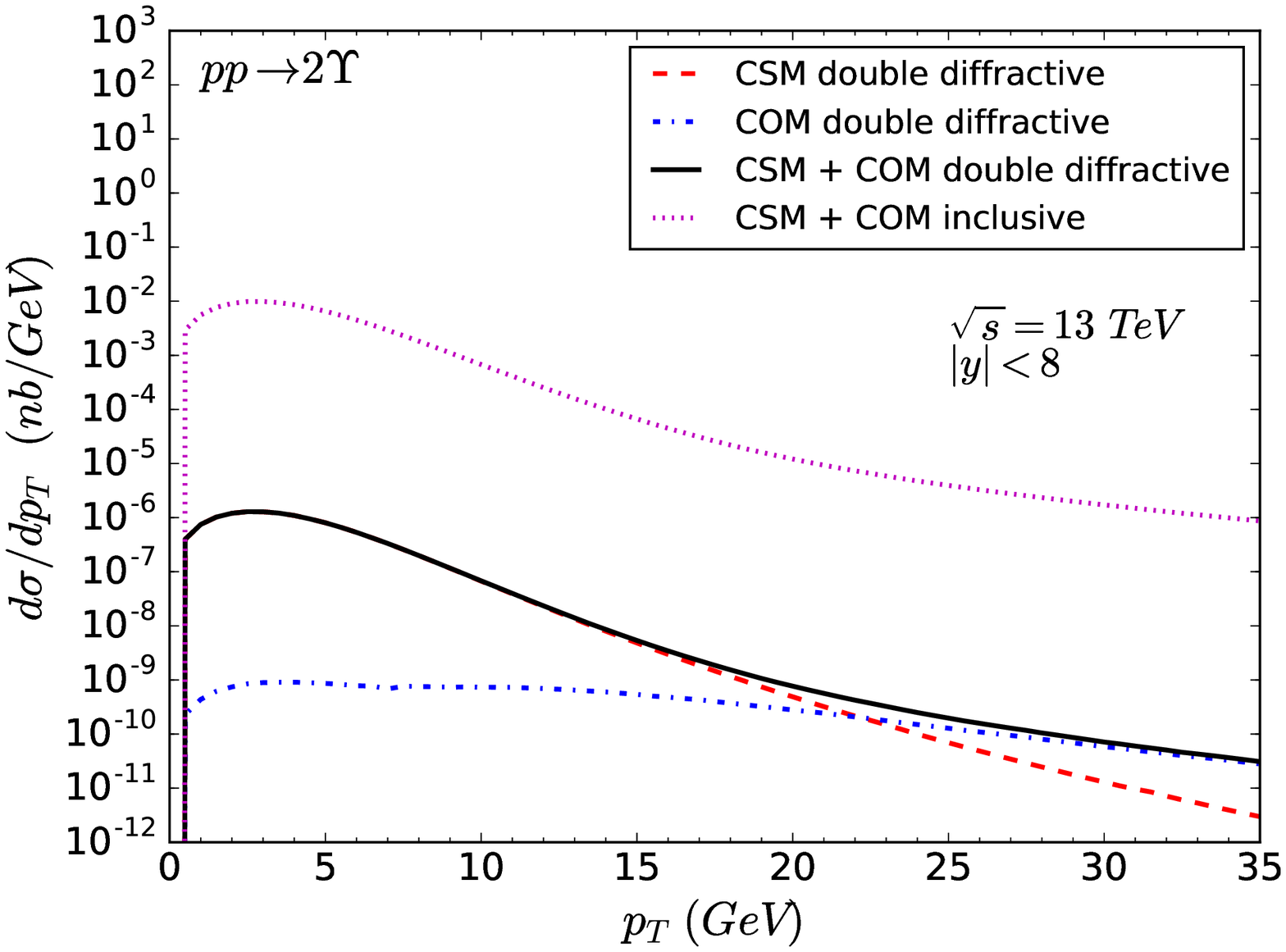}
}}
\end{tabular}
\caption{ $p_T$ distribution for double $J/\Psi$ production (left) and double $\Upsilon$ production (right) in double diffractive process.} 
\label{DD}
\end{figure*}

For the gluon-initiated color singlet contributions (CSM), $\frac{d\hat{\sigma}}{d\hat{t}}$ is calculated in Ref. \cite{Qiao}. One important parameter is the quarkonium squared radial function at the origin $|R(0)|^2$, which is related to the leptonic decay rate \cite{Eichten} as 
$\Gamma(H\to e^+e^-)=\frac{4N_c\alpha^2e_Q^2}{3}\frac{|R(0)|^2}{M_H^2}\left(1-\frac{16\alpha_s}{3\pi}\right)$\,. From recent PDG data \cite{pdg} 
for $\Gamma(J/\Psi\to e^+e^-)$ and $\Gamma(\Upsilon \to e^+e^-)$, we obtain $|R_{J/\Psi}(0)|^2=0.53\,$ GeV$^3$ and $|R_{\Upsilon}(0)|^2=4.6 \,$ GeV$^3$ for the $J/\Psi$ and the $\Upsilon$, respectivelly. For the charm and bottom quark masses we use $m_c=1.5\,$ GeV and $m_b=4.7\,$ GeV, respectivelly.

For the color octet contributions, the differential cross section $\frac{d\hat{\sigma}}{d\hat{t}}$ for the gluon initiated partonic subprocesses is calculated in Ref. \cite{Ko}. In this case, the only relevant NRQCD matrix elements are $\langle O_8^{J/\psi}({}^3S_1) \rangle=3.9\times 10^{-3} GeV^3$ and $\langle O_8^{\Upsilon}({}^3S_1) \rangle=1.5\times 10^{-1} GeV^3$, taken from \cite{Braaten2000}.
For the $J/\Psi$, this value has been updated in a recent fit to world data \cite{bute} which gives $\langle O_8^{J/\psi}({}^3S_1) \rangle=1.68\times 10^{-3} GeV^3$. Using this new value, our results for the COM contributions would decrease by a factor $2.3$, which is inside the uncertainties of our results.

\begin{table*}
\centering
\begin{tabular}{|l|c|c|c|}
\hline 
~ & {\bf Inclusive} & {\bf Single Diffrative} & {\bf Double Diffractive} \\ 
\hline
\hline 
$\sigma^{J/\Psi J/\Psi}$ & 28,3 nb & $3.8\times 10^{-1}$ nb & $8.8\times 10^{-3}$ nb \\ 
\hline 
$\sigma^{\Upsilon \Upsilon}$ & 52.3 pb & $4.5\times 10^{-1}$ pb & $6.6\times 10^{-3}$ pb \\ 
\hline
\hline 
$\sigma^{J/\Psi J/\Psi}$ (LHCb) & 6.04 nb & $7.8\times 10^{-2}$ nb & $1.7\times 10^{-3}$ nb \\ 
\hline 
$\sigma^{\Upsilon \Upsilon}$ (LHCb) & 10.4 pb & $8.2\times 10^{-2}$ pb & $8.26\times 10^{-4}$ pb \\ 
\hline 
\end{tabular}
\caption{The total cross sections of the single and double diffractive production of $J/\Psi J/\Psi$ and $\Upsilon \Upsilon$.}
\label{tabela}
\end{table*}

\section{Results}

In what follows we show our predictions for the production of $J/\Psi J/\Psi$ and $\Upsilon \Upsilon$ in $pp$ collisions at the Run 2 LHC ($\sqrt{s}=13\, TeV$) energies, which complement the predictions presented in Ref. \cite{Mariotto}.

In figure \ref{Rapidez} we present our predictions for the differential cross section distribution in rapidity for double $J/\Psi$ and double $\Upsilon$. The inclusive process is also presented for comparison. The rapidity distributions are flat at the central region ($y \approx 0$) and the results are dominated by color singlet contribution. We have that for central rapidities, the inclusive prediction is a factor $10$ ($10^3$) larger than SD (DD) prediction for $J/\Psi J/\Psi$ production. For $\Upsilon \Upsilon$ the inclusive prediction is a factor $10^2$ ($10^4$) larger than SD (DD) prediction.

Our results for the $p_T$ distributions associated to the single and double diffractive processes, are shown in figures \ref{SD} and \ref{DD}, respectively. In the case of SD processes, one have that the Color Octet (COM) contribution becomes dominant for $p_T \ge 10$ GeV for $J/\Psi J/\Psi$ and  $p_T \ge 20$ GeV for $\Upsilon \Upsilon$. Similar behavior is also present in the case of the DD processes. As we can see in the $p_T$ distributions, the CSM is the most important mechanism for the low $p_T$ region, being the large $p_T$ tail dominated by the COM contributions. For the $p_T$-integrated rapidity cross section, only the CSM contributes.

In table \ref{tabela} we also show our predictions for the total cross section ($\sqrt{s} = 13$ TeV) in different rapidity ranges. In the first two lines we consider the rapidity range $|y|<8$. The following lines present the estimates for the rapidity range probed by the LHCb detector ($2 < y < 4.5$).

\section{Conclusions}
In this work we calculate the production of a $J/\Psi$ pair and a $\Upsilon$ pair via hard single and double diffractive processes at LHC $\sqrt{s} = 13$ TeV energies. Considering the NRQCD formalism we estimate the differential and total cross sections for these processes. Our results demonstrate that the contribution of diffractive processes have non negligible orders of magnitude, and therefore this study can be useful to test the Pomeron structure. With respect to the inclusive production our results demonstrate that the contribution of color singlet and color octet processes are important in different kinematic regions, and that these processes are feasible at the LHC.

\begin{acknowledgments}
This work was partially financed by the Brazilian funding agencies CAPES, CNPq, FAPERGS and INCT-FNA (process number 464898/2014-5).
\end{acknowledgments}

\hspace{1.0cm}

\end{document}